\documentclass[aps,showpacs,preprintnumbers,amsmath,nofootinbib]{revtex4}
\usepackage{amssymb}
\usepackage{amsmath}
\usepackage{graphics}
\usepackage{graphicx}
\usepackage{dcolumn}
\usepackage{bm}
\usepackage{color}
\begin{document}

\newcommand{\blue}[1]{\textcolor{blue}{#1}}
\newcommand{\new}{\blue}
\newcommand{\green}[1]{\textcolor{green}{#1}}
\newcommand{\modif}{\green}
\newcommand{\red}[1]{\textcolor{red}{#1}}
\newcommand{\attention}{\red}

\title{The energy dependence of the ratio of elastic to total cross section\\
in $pp$ and $\bar{p}p$ collisions}

\author{V.A. Okorokov} \email{VAOkorokov@mephi.ru; Okorokov@bnl.gov}
\affiliation{National Research
Nuclear University MEPhI (Moscow Engineering Physics Institute), \\
Kashirskoe highway 31, 115409 Moscow, Russia}

\date{\today}

\begin{abstract}
The paper presents the phenomenological analysis of the energy dependence for
the ratio of elastic to total cross section in proton-proton and antiproton-proton scattering. The analytic functions based on the study of low- and high-energy experimental data for various scattering parameters provide the quantitative description of energy dependence of the ratio with statistically acceptable qualities in wide range of collision energy $\sqrt{s} \geq 3$ GeV in the case of the separate datasets for $pp$, $\bar{p}p$ collisions and at $\sqrt{s} \geq 5$ GeV for the joined experimental data ensemble. Based on the fit results the estimations are derived for the ratio of elastic to total cross section in $pp$ scattering at various $\sqrt{s}$ up to energy frontier $\sqrt{s}=10$ PeV which can be useful for present and future hadron colliders as well as for cosmic ray measurements at ultra-high energies. The indication is observed for onset of the asymptotic region at $\sqrt{s} \gtrsim 5-10$ PeV for the ratio of cross sections under consideration.

\end{abstract}


\maketitle

%%%%%%%%%%%%%%%%%%%%%%%%%%%%%%%%%%%%%%%%%%%%%%%%%%%%%%%%%%%%%%%%%%%%%%%%
\section{Introduction}\label{sec:1}

Experimental data show definitely that the elastic cross section ($\sigma_{\scriptsize{\mbox{el}}}$) growth faster with increasing of collision energy ($\sqrt{s}$) than total cross section ($\sigma_{\scriptsize{\mbox{tot}}}$) in $pp$ and $\bar{p}p$ collisions. This means that the probability for survival of incoming (anti)nucleons increases with the $\sqrt{s}$ and this experimental fact is important and some unexpected. It should be stressed the nature of the effect is yet not understood, moreover, the advancing growth of the $\sigma_{\scriptsize{\mbox{el}}}(s)$ with respect to the $\sigma_{\scriptsize{\mbox{tot}}}(s)$ is studied much less intensively than, for example, the behavior of the energy dependence of the total cross section for $pp$ and $\bar{p}p$ collisions. The possible interrelations the energy behavior of $\sigma_{\scriptsize{\mbox{el}}}$ and, as consequence, survival probability in (anti)nucleon collisions with Bose--Einstein condensation and topology of the QCD vacuum are considered in \cite{PU-188-437-2018}. As discussed in \cite{Okorokov-UJPA-1-196-2013} the structure of the QCD vacuum may be characterized by highly irregular, fractal-like geometry. Furthermore one can suggest that the geometric shape of interaction region may be complex and correspond rather fractal than the simplest cases of the Euclidean geometry. Thus the hypothesis can be suggested for influence both the topology properties of the QCD vacuum and the type of geometry used for approximation for the shape of colliding particles on energy dependence of global parameters, especially, cross sections of hadronic scattering. In general this suggestion means the fundamental interrelation between geometry and dynamics of the strong interaction processes. But these qualitative reasoning should be verified by analysis of experimental data.

Another important task is the study of asymptotic behavior of the global parameters of hadronic scattering and search for the onset for asymptotic region on collision energy. Asymptotic relations are based on the fundamental properties of the quantum field theory, namely, on the analyticity of scattering amplitude and unitarity condition. Moreover processes with small momentum transfer provide the main contribution in cross sections of hadronic interactions at high energies. Thus the investigations of the energy behavior of cross sections and relations between them allow the wide opportunities for study of QCD-inspired models, in particular, the approach of gluonic (or glueballic) clouds, on large distances.

Therefore the study of energy dependence for relation between $\sigma_{\scriptsize{\mbox{tot}}}$ and $\sigma_{\scriptsize{\mbox{el}}}$ in hadronic interactions seems relevant for development both the theory of strong interactions, in particular, soft physics and the basic statements of quantum field theory.

The paper is organized as follows. In Section \ref{sec:2}, the definition of observable and equations for fitting functions are presented. Section \ref{sec:3} contains the results for fits of the ratio of elastic to total cross section in $pp$ and $\bar{p}p$ elastic scattering separately as well as in the case of joined sample for these two types of collisions. In Section \ref{sec:4} the predictions for the cross section ratio in $pp$ at some energies as well as possible indication on the onset of the asymptotic region are discussed. Section \ref{sec:5} contains the final remarks and brief summary.

%%%%%%%%%%%%%%%%%%%%%%%%%%%%%%%%%%%%%%%%%%%%%%%%%%%%%%%%%%%%%%%%%%%%%%%%
\section{Observable and approximating functions}\label{sec:2}

In the present work the ratio of the elastic cross section to the total cross section
\begin{equation}
R_{\scriptsize{\mbox{e/t}}} = \sigma_{\scriptsize{\mbox{el}}} /
\sigma_{\scriptsize{\mbox{tot}}} \label{eq:2.1}
\end{equation}
is studied in dependence on the collision energy for $pp$ and
$\bar{p}p$ collisions. For the first case the experimental sample
for set of the global scattering parameters $\mathcal{G}_{pp} \equiv
\{\mathcal{G}_{pp}^{i}\}_{i=1}^{2}=\{\sigma_{\footnotesize\mbox{tot}}^{pp},
\sigma_{\footnotesize\mbox{el}}^{pp}\}$ based on the database from
\cite{Okorokov-IJMPA-33-1850077-2018}, for $\bar{p}p$ scattering
experimental points for the corresponding set
$\mathcal{G}_{\bar{p}p}$ are from
\cite{Okorokov-IJMPA-32-1750175-2017}.

The $\sigma_{\footnotesize\mbox{tot}}(s)$ and $\sigma_{\footnotesize\mbox{el}}(s)$ demonstrate similar basic features and functional behavior in wide energy range. In general the behavior of the $\sigma_{\footnotesize\mbox{tot}}$ in high energy domain can be qualitatively described as $\propto \ln^{\alpha}\varepsilon$ for $pp$ and $\bar{p}p$ collisions, where $0 < \alpha \leq 2$, $\varepsilon \equiv s/s_{0}$ and $s_{0}=1$ GeV$^{2}$. Such dependence appears in various approaches \cite{Okorokov-IJMPA-33-1850077-2018,Okorokov-IJMPA-32-1750175-2017,PDG-ChP-C40-100001-2016}.
On the other hand the term $\propto \varepsilon^{-\beta}$ suggested within axiomatic quantum field theory (AQFT) \cite{Okorokov-IJMPA-25-5333-2010} allows us the reasonable description for the experimental data at low energies down to the $\sqrt{s}=3$ GeV \cite{Okorokov-IJMPA-32-1750175-2017,Okorokov-IJMPA-25-5333-2010}. Moreover the detailed analysis of the energy dependence of slope parameter ($B$) in elastic scattering shows that the function $\propto \ln^{-\gamma}\varepsilon$ provides the close $\chi^{2}/\mbox{n.d.f.}$ values as the power law for low-energy domain \cite{Okorokov-AHEP-2015-914170-2015}. Therefore both functions under discussion can be considered as an possible approximation for the low-energy behavior of parameters from the set $\mathcal{G}_{pp/\bar{p}p}$. Taking into account the arguments given above as well as the view of experimental dependence for the ratio (\ref{eq:2.1}) shown below the following analytic functions are suggested for approximation of the data for
$R_{\scriptsize{\mbox{e/t}}}(s)$:
\begin{subequations}
\label{eq:2.2}
\begin{eqnarray}
R_{\scriptsize{\mbox{e/t}}}(s)&=& a_{1}+a_{2}\ln^{a_{3}}
\varepsilon +a_{4}\ln^{-a_{5}} \varepsilon,
\label{eq:2.2a} \\
R_{\scriptsize{\mbox{e/t}}}(s)&=& a_{1}+a_{2}\ln^{a_{3}}
\varepsilon +a_{4}\varepsilon^{-a_{5}}, \label{eq:2.2b}
\end{eqnarray}
\end{subequations}
where free parameters $a_{i}$, $i=1-5$ depend on range of the fit, i.e. on
the low boundary for energy interval $s \geq
s_{\scriptsize{\mbox{min}}}$. It is supposed the second terms in
the functions (\ref{eq:2.2}) describe the high-energy behavior of
$R_{\scriptsize{\mbox{e/t}}}(s)$ and the third terms -- low-energy
domain.

It was deduced, that upper bound on the inelastic cross-section ($\sigma_{\scriptsize{\mbox{inel}}}$) at high energy is one-fourth
of the corresponding upper bound on $\sigma_{\scriptsize{\mbox{tot}}}$ \cite{PRD-80-065015-2009}, i.e. $\bigl[\sup\sigma_{\scriptsize{\mbox{inel}}}(s)/\sup\sigma_{\scriptsize{\mbox{tot}}}(s)\bigr]_{s \in\,\mathbf{S_{h}}\,\subset\,\mathbf{S}}=1/4$ in the high energy subrange $\mathbf{S_{h}} \subset \mathbf{S}$ with taken into account the equal domain of energies $\mathbf{S} \equiv [4m_{p}^{2}, \infty)$ on which any cross section ($\sigma_{\scriptsize{\mbox{tot}}}$, $\sigma_{\scriptsize{\mbox{el}}}$ and $\sigma_{\scriptsize{\mbox{inel}}}$) is defined, where $m_{p}$ is the proton mass \cite{PDG-ChP-C40-100001-2016} and the low boundary of $\mathbf{S}$ -- $s_{\scriptsize{\mbox{l.b.}}} \equiv 4m_{p}^{2}$ -- corresponds to the interactions ($pp$, $\bar{p}p$) under discussion. According to the propositions for the properties of scattering amplitude for binary process $\varkappa_{1} + \varkappa_{2} \to \varkappa_{3} + \varkappa_{4}$ within AQFT the cross sections $\sigma_{\scriptsize{\mbox{tot}}}(s)$, $\sigma_{\scriptsize{\mbox{el}}}(s)$ are locally integrable functions at $s \in \mathbf{S}$ \cite{Bogolubov-book-1990}. It means that these functions are Riemann integrable $\forall\,[s_{1},s_{2}] \subset \mathbf{S}$. In view of the necessary condition for Riemann integrability of a function, the cross sections $\sigma_{\scriptsize{\mbox{tot}}}(s)$, $\sigma_{\scriptsize{\mbox{el}}}(s)$ are bounded $\forall\,[s_{1},s_{2}] \subset \mathbf{S}$ (see, for instance, \cite{Zorich-book-2002}). All these properties are valid for $\sigma_{\scriptsize{\mbox{inel}}}(s)$ due to the unitarity condition
$\sigma_{\scriptsize{\mbox{tot}}}=\sigma_{\scriptsize{\mbox{el}}}+\sigma_{\scriptsize{\mbox{inel}}}$ and the linearity property. It is considered reliably established, both theoretically and experimentally, that all the cross sections under consideration grow with increasing $s$ at least for high energy domain \cite{PDG-ChP-C40-100001-2016}. Therefore the following relations are valid: (i) $\forall\,i$, $i$ = tot, inel, el: $\exists\,\lim_{s \to \infty} \sigma_{i}(s) \equiv L_{i}$ and $L_{i}$ is finite; (ii) $\forall\,i$, $i$ = tot, inel, el: $\sup_{s\,\in\,\mathbf{S_{h}}} \sigma_{i}(s) = \lim_{s \to \infty} \sigma_{i}(s)$. Then
\begin{equation}
\displaystyle
\frac{\sup_{s\,\in\,\mathbf{S_{h}}\,\subset\,\mathbf{S}}\sigma_{\scriptsize{\mbox{inel}}}(s)}
{\sup_{s\,\in\,\mathbf{S_{h}}\,\subset\,\mathbf{S}}\sigma_{\scriptsize{\mbox{tot}}}(s)}=
\frac{\lim_{s \to \infty}\sigma_{\scriptsize{\mbox{inel}}}(s)}
{\lim_{s \to \infty}\sigma_{\scriptsize{\mbox{tot}}}(s)}=\lim_{s \to \infty}\frac{\sigma_{\scriptsize{\mbox{inel}}}(s)}{\sigma_{\scriptsize{\mbox{tot}}}(s)} \equiv \lim_{s \to \infty}R_{\scriptsize{\mbox{i/t}}}(s),
\label{eq:2.3add}
\end{equation}
where $R_{\scriptsize{\mbox{i/t}}} = \sigma_{\scriptsize{\mbox{inel}}} /
\sigma_{\scriptsize{\mbox{tot}}}$. It means
\begin{equation}
\bigl[R_{\scriptsize{\mbox{e/t}}}(s)]_{s \to \infty} \to 3/4 \label{eq:2.3}
\end{equation}
with taking into account the unitarity condition. Asymptotically both approximating functions (\ref{eq:2.2}) can be written as follows:
\begin{equation}
\bigl[R_{\scriptsize{\mbox{e/t}}}(s)]_{s \to \infty} \approx a_{1}+a_{2}\ln^{a_{3}}
\varepsilon \label{eq:2.4}
\end{equation}
because the terms corresponding to the low-energy region become negligibly small at $s \to \infty$ in the Eqs. (\ref{eq:2.2a}) and (\ref{eq:2.2b}). Consequently, the following relation is derived
\begin{equation}
%\displaystyle
%s_{\scriptsize{\mbox{a}}} \simeq s_{0}\exp\biggl[\biggl(\frac{3/4-a_{1}}{a_{2}}\biggr)^{1/a_{3}}\biggr] \label{eq:2.5}
s_{\scriptsize{\mbox{a}}} \sim s_{0}\exp\left\{\bigl[(3/4-a_{1})/a_{2}\bigr]^{1/a_{3}}\right\} \label{eq:2.5}
\end{equation}
for the collision energy at which asymptotic regime should manifests itself in the $pp$ and $\bar{p}p$ interactions\footnote{It is obvious that the energy scale of the boundary of the asymptotic region is a physically objective quantity that is constant, at least for the certain indication under consideration, and does not depend on the specific form of approximating function for this indication. The view of this function determines only the certain relation between the estimation of the asymptotic energy and the fit parameters.}.

Also the quantity (\ref{eq:2.1}) can be related with the parameter $\kappa(s) \equiv \sigma_{\footnotesize\mbox{tot}}(s)/8\pi B(s)$ indicating the asymptotic domain for such global scattering parameters like total cross sections, slope etc. \cite{Okorokov-IJMPA-33-1850077-2018} as follows
\begin{equation}
\kappa(s)=2R_{\scriptsize{\mbox{e/t}}}(s)
\bigl[1+\rho^{2}(s)\bigr]^{-1},~~~~~ \lim_{\rho \to 0}\kappa(s)=2R_{\scriptsize{\mbox{e/t}}}(s), \label{eq:2.6}
\end{equation}
where $\rho(s) \equiv \mathrm{Re}F(s,0)/\mathrm{Im}F(s,0)$ is the ratio of real
to imaginary part of the amplitude in the forward direction. One can note that experimental data show $|\rho(s)| \lesssim
0.3\,(0.2)$ at $\sqrt{s} \gtrsim 5\,(3)$ GeV for $pp$ ($\bar{p}p$)
collisions \cite{Okorokov-IJMPA-32-1750175-2017}. Therefore the limit in (\ref{eq:2.6}) is valid in wide energy ranges for both $pp$ and $\bar{p}p$ scattering at accuracy level better than 0.1.
%%%%%%%%%%%%%%%%%%%%%%%%%%%%%%%%%%%%%%%%%%%%%%%%%%%%%%%%%%%%%%%%%%%%%%%%
\section{Experimental data and fit results}\label{sec:3}

In Figs. \ref{fig:1} and \ref{fig:2} the dependence of the ratio
(\ref{eq:2.1}) on $\sqrt{s}$ is shown for $pp$ and $\bar{p}p$
collisions respectively\footnote{In the paper total errors are
used for experimental points unless otherwise specified. The total
error is calculated as addition of systematic and statistical
uncertainties in quadrature \cite{PDG-ChP-C40-100001-2016}.}.
The physical observable $R_{\scriptsize{\mbox{e/t}}}$ is
experimentally available for $\sqrt{s} \geq 2.03$ (1.89) GeV in
$pp$ ($\bar{p}p$) elastic scattering. As seen there is the sharp
peak in the $R_{\scriptsize{\mbox{e/t}}}(s)$ near the low
threshold $2m_{p}$ for $\bar{p}p$ scattering (Fig. \ref{fig:2})
which does not allow us to use the equations (\ref{eq:2.2}) for
approximation of all available experimental points. Therefore a
low boundary of the energy range for approximations with help of
the functions (\ref{eq:2.2}) should be chosen no smaller than 2
GeV. Furthermore in Fig. \ref{fig:1} the $R_{\scriptsize{\mbox{e/t}}}(s)$ qualitatively
agrees with constant at $\sqrt{s} \geq 100$ GeV and possibly this
statement is valid for $\bar{p}p$ too (Fig. \ref{fig:2}). Thus the
dependence $R_{\scriptsize{\mbox{e/t}}}(s)$ is also fitted by
constant $R_{\scriptsize{\mbox{e/t}}}(s)=a_{0}$ at $\sqrt{s_{\scriptsize{\mbox{min}}}}=100$
GeV for both the $pp$ and $\bar{p}p$ elastic scattering reactions\footnote{As seen in Figs. \ref{fig:1}, \ref{fig:2} there is large gap between measurements at the highest Intersecting Storage Rings (ISR) energy $\sqrt{s} \approx 63$ GeV and nearest right point at $\sqrt{s}=2760$ (546) GeV for $pp$ ($\bar{p}p$). Therefore the choice $\sqrt{s}=100$ GeV as some boundary is conditional and this value is used below without loss of generality}.
Tables \ref{tab:1}, \ref{tab:2} show the fit parameter values for
approximations of the experimental
$R_{\scriptsize{\mbox{e/t}}}(s)$ for $pp$ and $\bar{p}p$
respectively by functions under consideration at various
$s_{\scriptsize{\mbox{min}}}$.

At $\sqrt{s_{\scriptsize{\mbox{min}}}}=2$ GeV function (\ref{eq:2.2a}) provides significantly better $\chi^{2}/\mbox{n.d.f.}$ than function (\ref{eq:2.2b}) but
the both approximations (\ref{eq:2.2}) under consideration lead to the poor fit quality in the case of elastic $pp$ interactions with opposite situation for $\bar{p}p$ with regard of fit qualities for (\ref{eq:2.2a}) and (\ref{eq:2.2b}). From Tables \ref{tab:1}, \ref{tab:2} it is seen that the both functions (\ref{eq:2.2}) approximate the experimental dependence $R_{\scriptsize{\mbox{e/t}}}(s)$ for $pp$, $\bar{p}p$ with statistically acceptable quality for wide energy range $\sqrt{s} \geq 3$ GeV, which covers the domain of validity of the approach $1+\rho^{2} \approx 1$ at accuracy level not worse than 0.09, 0.04 respectively. Therefore in the rest part of paragraph the discussion is focused on the results obtained at $\sqrt{s_{\scriptsize{\mbox{min}}}} \geq 3$ GeV (Tables \ref{tab:1}, \ref{tab:2}). For the $pp$ collisions the behavior of the generalized logarithmic function (\ref{eq:2.2a}) is close to the $\ln \varepsilon$ at high energies, especially for $\sqrt{s_{\scriptsize{\mbox{min}}}}=3$ GeV while the function (\ref{eq:2.2b}) leads to the significantly faster growth in high-energy range at corresponding $s_{\scriptsize{\mbox{min}}}$, namely, the $a_{3}$ value almost coincides with 2.0 within 2 standard deviation (s.d.). In the case of $\bar{p}p$ the $a_{3}$ is characterized by larger dispersion of values (Table \ref{tab:2}): the function (\ref{eq:2.2a}) is close to the $\ln \varepsilon$ at high energies for $\sqrt{s_{\scriptsize{\mbox{min}}}}=3$ GeV only while $a_{3}$ value coincides with 2.0 within 2 s.d. at larger $s_{\scriptsize{\mbox{min}}}$; the function (\ref{eq:2.2b}) provides large discrepancy for different $s_{\scriptsize{\mbox{min}}}$ at high energies, in particular, the (\ref{eq:2.2b}) growths very fast at $\sqrt{s_{\scriptsize{\mbox{min}}}}=3$ GeV. Although the functions (\ref{eq:2.2a}) and (\ref{eq:2.2b}) show close values of $\chi^{2}/\mbox{n.d.f.}$ the behavior of $R_{\scriptsize{\mbox{e/t}}}(s)$ for small and intermediate energies is slightly better described by the power law dependence on $\varepsilon$ than the power of $\ln \varepsilon$. At $\sqrt{s_{\scriptsize{\mbox{min}}}}=100$ GeV constant provides the quantitative agreement with experimental points for both types of collisions ($pp$ and $\bar{p}p$).

The comparison of fitting curves obtained for various functions (\ref{eq:2.2}) and values of $s_{\scriptsize{\mbox{min}}}$ provides the following qualitative conclusions. In accordance with numerical fit results (Tables \ref{tab:1}, \ref{tab:2}) the dependence on the type of fitting function at fixed value $\sqrt{s_{\scriptsize{\mbox{min}}}}=2-5$ GeV is most pronounced for the smallest value of the low boundary for the fitted interval of $s$ and becomes weaker with increasing of the $s_{\scriptsize{\mbox{min}}}$ for both $pp$ and $\bar{p}p$ collisions. For $\sqrt{s_{\scriptsize{\mbox{min}}}}=2$ GeV visible difference is observed between curves obtained with help of the functions (\ref{eq:2.2a}) and (\ref{eq:2.2b}) at $\sqrt{s} \gtrsim 100$ GeV for $pp$ and even at $\sqrt{s} \gtrsim 20$ GeV for $\bar{p}p$. The range of close behavior for (\ref{eq:2.2a}) and (\ref{eq:2.2b}) significantly expands for larger $s_{\scriptsize{\mbox{min}}}$. The fit curves are close to each other for $\sqrt{s} \gtrsim 10$ (1) TeV at $\sqrt{s_{\scriptsize{\mbox{min}}}}=3$ GeV and for $\sqrt{s} \gtrsim 100$ (10) TeV at $\sqrt{s_{\scriptsize{\mbox{min}}}}=5$ GeV in $pp$ ($\bar{p}p$) scattering, i.e. in the last case the fit curve for proton-proton or antiproton-proton data depends on type of functions (\ref{eq:2.2}) very weakly at any experimentally reached energies.
The situations with fitting curves at fixed type of function in (\ref{eq:2.2}) and variation of the $s_{\scriptsize{\mbox{min}}}$ are similar for both type of collisions under study. The fitting curves (\ref{eq:2.2a}) show a close behavior for all $s_{\scriptsize{\mbox{min}}}$, especially at $\sqrt{s_{\scriptsize{\mbox{min}}}}=3$ and 5 GeV in the experimentally measured energy domain with some larger discrepancy between approximations for $\bar{p}p$ than that for $pp$. For the function (\ref{eq:2.2b}) the curve for $\sqrt{s_{\scriptsize{\mbox{min}}}}=2$ GeV differs noticeably from the other two approximations, which practically coincide with each other at $5 \leq \sqrt{s} < 10^{5}$ (2000) GeV for $pp$ ($\bar{p}p$). The qualitative relations indicated above for various fitting curves are mostly explained by influence of the low-energy data which are excluded from the fitted samples at $\sqrt{s_{\scriptsize{\mbox{min}}}} \geq 3$ GeV.

In Figs. \ref{fig:1}, \ref{fig:2} the smooth curves correspond to the fits with best quality at each certain $\sqrt{s_{\scriptsize{\mbox{min}}}}=2$ and 3 GeV, i.e. the fit by function (\ref{eq:2.2a}) at $\sqrt{s_{\scriptsize{\mbox{min}}}}=2$ GeV and by (\ref{eq:2.2b}) at $\sqrt{s_{\scriptsize{\mbox{min}}}}=3$ GeV are shown for $pp$ collisions; the approximations by the function (\ref{eq:2.2b}) for both $\sqrt{s_{\scriptsize{\mbox{min}}}}$ under consideration are presented for $\bar{p}p$ elastic reaction in accordance with Tables \ref{tab:1}, \ref{tab:2}. As seen in Figs. \ref{fig:1}, \ref{fig:2} the suggested functions (\ref{eq:2.2a}) and (\ref{eq:2.2b}) provide the reasonable agreement with experimental data down to the $\sqrt{s_{\scriptsize{\mbox{min}}}}=2$ GeV for $pp$ (Fig. \ref{fig:1}) and $\bar{p}p$ scattering (Fig. \ref{fig:2}) respectively. The fit curves shown in Fig. \ref{fig:1} for $pp$ demonstrate the close behavior at all experimentally available energies while for $\bar{p}p$ the approximation at $\sqrt{s_{\scriptsize{\mbox{min}}}}=2$ GeV is characterized significantly slower growth in high-energy range than the curve at $\sqrt{s_{\scriptsize{\mbox{min}}}}=3$ GeV (Fig. \ref{fig:2}). The inner panel in Fig. \ref{fig:2} shows the change of trend in $R_{\scriptsize{\mbox{e/t}}}(s)$ near the low-boundary energy $2m_{p}$ in details for $\bar{p}p$ collisions and confirms the validity of approximations (\ref{eq:2.2}) at $\sqrt{s} \geq 2$ GeV. As indicated above the fit curve (\ref{eq:2.2b}) at $\sqrt{s_{\scriptsize{\mbox{min}}}}=3$ GeV for $\bar{p}p$ demonstrate very fast growth for multi-TeV region in comparison with both the other approximations for $\bar{p}p$ and the corresponding curve for $pp$. But this difference can be dominated by absent of experimental points for the first case at $\sqrt{s} > 2$ TeV. Therefore the new data for $\bar{p}p$ are crucially important at multi-TeV energies for more definite physical conclusions with regard of high-energy behavior of $R_{\scriptsize{\mbox{e/t}}}(s)$ for $\bar{p}p$. Results of the present work are compared with the ratio of approximation for $\sigma_{\footnotesize\mbox{el}}$ from \cite{TOTEM-arXiv-1712.06153} to "standard" functions for $\sigma_{\footnotesize\mbox{tot}}$ in $pp$ and $\bar{p}p$ reactions from \cite{PDG-ChP-C40-100001-2016} shown by thin lines in Figs. \ref{fig:1}, \ref{fig:2}. As expected these curves lie some lower than most of experimental points at $\sqrt{s} < 10$ GeV because the fit for $\sigma_{\footnotesize\mbox{el}}$ was only made for $\sqrt{s} \geq 10$ GeV  \cite{TOTEM-arXiv-1712.06153}. For $pp$ collisions there is the visible discrepancy between smooth curves (\ref{eq:2.2}) and approximation based on the parameterizations from \cite{TOTEM-arXiv-1712.06153,PDG-ChP-C40-100001-2016} in the energy domain $\sqrt{s} \sim 0.3 - 30$ TeV (Fig. \ref{fig:1}). The new experimental data from Relativistic Heavy Ion Collider (RHIC) at $\sqrt{s} \sim 0.1 - 0.5$ TeV and from low-energy Large Hadron Collider (LHC) mode at $\sqrt{s} \sim 1.0$ TeV will be helpful for choice of favorable approach. The curve for $pp$ based on the \cite{TOTEM-arXiv-1712.06153,PDG-ChP-C40-100001-2016} shows the slower growth of $R_{\scriptsize{\mbox{e/t}}}(s)$ at ultra-high energies $\sqrt{s} \gtrsim 100$ TeV with regard of functions (\ref{eq:2.2}) at various $s_{\scriptsize{\mbox{min}}}$ and this feature can be important for estimation of the onset of asymptotic regime. The approximating line for $\bar{p}p$ obtained with help of the parameterizations from \cite{TOTEM-arXiv-1712.06153,PDG-ChP-C40-100001-2016} is very close to the curves with best fit qualities obtained within present work at $\sqrt{s_{\scriptsize{\mbox{min}}}}=2$ and 3 GeV in the energy range $10 \lesssim \sqrt{s} \lesssim 2000$ GeV then the some discrepancy appears for higher energies (Fig. \ref{fig:2}). This difference between various approximations for $R_{\scriptsize{\mbox{e/t}}}(s)$ in $\bar{p}p$ increases with growth of $s$.

As seen in Figs. \ref{fig:1}, \ref{fig:2} the experimental dependence $R_{\scriptsize{\mbox{e/t}}}(s)$ shows the similar functional behavior at qualitative level for $pp$ and $\bar{p}p$ collisions at $\sqrt{s} > 2$ GeV, i.e. in full energy domain which is available for comparison. Therefore the ratio (\ref{eq:2.1}) is studied for joined sample for $pp$ and $\bar{p}p$ collisions\footnote{Below for brevity the joined sample is also called the ensemble for nucleon-nucleon scattering.}. Fig. \ref{fig:3} shows the experimental dependence $R_{\scriptsize{\mbox{e/t}}}(s)$ together with approximations described in details below for nucleon-nucleon scattering. Numerical results of fitting the $R_{\scriptsize{\mbox{e/t}}}(s)$ by functions (\ref{eq:2.2}) are shown in Table \ref{tab:3} at various $\sqrt{s_{\scriptsize{\mbox{min}}}}$. Functions (\ref{eq:2.2a}) and (\ref{eq:2.2b}) approximate the joined dataset for the ratio (\ref{eq:2.1}) in nucleon-nucleon scattering with close values of $\chi^{2}/\mbox{n.d.f.}$ at any $\sqrt{s_{\scriptsize{\mbox{min}}}}=2-5$ GeV and statistically acceptable quality is achieved at $\sqrt{s_{\scriptsize{\mbox{min}}}}=5$ GeV only. Moreover the function (\ref{eq:2.2b}) provides the some better quality of the approximation especially at $\sqrt{s_{\scriptsize{\mbox{min}}}}=2$ GeV than the generalized logarithmic function (\ref{eq:2.2a}). In difference with the fits for separated samples for $pp$ and $\bar{p}p$ approximation of the $R_{\scriptsize{\mbox{e/t}}}(s)$ for joined ensemble for nucleon-nucleon collisions by a constant with statistically acceptable quality is possible only in the TeV-energy range. Since the main contribution is given by experiments on $pp$ collisions in this energy domain, then, as might be expected, the value of the $a_{0}$ for the joined sample coincides within errors with the value of the constant for $pp$ scattering. The inner panel in Fig. \ref{fig:3} shows the $R_{\scriptsize{\mbox{e/t}}}(s)$ near the low-boundary energy $2m_{p}$ in details for $pp$ and $\bar{p}p$ collisions. As seen this energy range is characterized by the significant dispersion of experimental points in $pp$, $\bar{p}p$ which leads to the unacceptable large $\chi^{2}/\mbox{n.d.f.}$ at $\sqrt{s_{\scriptsize{\mbox{min}}}}=2$ GeV. On the other hand the experimental points demonstrate the similar trends -- decreasing of the ratio (\ref{eq:2.1}) with the collision energy increase -- at $\sqrt{s} > 2.1-2.2$ GeV which lead to the fast improvement of the agreement between data for $pp$ and $\bar{p}p$. Thus the detailed analysis of low-energy data for nucleon-nucleon collisions confirms the qualitative observation made above with regard of the similar functional behavior of the $R_{\scriptsize{\mbox{e/t}}}(s)$ in $pp$ and $\bar{p}p$ scattering in almost whole energy domain which is available for comparison of experimental data for these two collision processes.

There are the following qualitative relations between fit curves for various parameterizations (\ref{eq:2.2}) and values of $s_{\scriptsize{\mbox{min}}}$. The fitting curves show a close behavior for all $s_{\scriptsize{\mbox{min}}}$ in the whole energy range achieved in the accelerator experiments in the case of the function (\ref{eq:2.2a}). Moreover this statement is valid for the parametrization (\ref{eq:2.2b}) too in difference with the separated samples for $pp$, $\bar{p}p$. For the nucleon-nucleon collisions the situation with respect to the influence of the type of parametrization at fixed $s_{\scriptsize{\mbox{min}}}$ is less unambiguous  than for $pp$ and $\bar{p}p$. For $\sqrt{s_{\scriptsize{\mbox{min}}}}=3$ and 5 GeV, visible difference between two curves (\ref{eq:2.2a}) and (\ref{eq:2.2b}) are observed at $\sqrt{s} \geq 10$ TeV, and in the first case the difference is much larger. But the approximations obtained with different functions in (\ref{eq:2.2}) are very close up to $\sqrt{s} \approx 100$ TeV and the difference between them becomes noticeable only at $\sqrt{s} \gtrsim 0.5$ PeV for the low boundary of the fitted energy domain $\sqrt{s_{\scriptsize{\mbox{min}}}}=2$ GeV.

During the pre-LHC period there was the hypothesis on constant behavior for quantity related with the $R_{\scriptsize{\mbox{e/t}}}(s)$ at intermediate energies \cite{Barone-book-2002}. Indeed Figs. \ref{fig:1} -- \ref{fig:3} show that the experimental $R_{\scriptsize{\mbox{e/t}}}(s)$ is almost flat in wide enough energy domain from $\sqrt{s} \sim 5-10$ GeV up to the $\sqrt{s} \approx 63$ GeV. For completeness the $R_{\scriptsize{\mbox{e/t}}}(s)$ is fitted by constant at intermediate energies $s_{1} \leq s \leq 10^{4}$ GeV$^{2}$ with the following results: $a_{0}=0.1789 \pm 0.0008$, $\chi^{2}/\mbox{n.d.f.}=28.3/28$ at $\sqrt{s_{1}}=10$ GeV for $pp$ collisions; $a_{0}=0.1805 \pm 0.0017$, $\chi^{2}/\mbox{n.d.f.}=13.2/15$ at $\sqrt{s_{1}}=5$ GeV for $\bar{p}p$ collisions; $a_{0}=0.1786 \pm 0.0008$, $\chi^{2}/\mbox{n.d.f.}=33.4/37$ at $\sqrt{s_{1}}=10$ GeV for nucleon-nucleon scattering. Approximation for joined ensemble is shown in Fig. \ref{fig:3} by thin line and numerical fit results confirms the hypothesis on constant behavior for $R_{\scriptsize{\mbox{e/t}}}(s)$ in both cases of the separate sets for $pp$, $\bar{p}p$ and of the joined sample for nucleon-nucleon scattering. But the measurements at TeV-energies unambiguously indicate that the hypothesis on constant $R_{\scriptsize{\mbox{e/t}}}(s)$ is valid in the some restricted energy domain and this effect is due to (i) wide and flat minimum in $\sigma_{\scriptsize{\mbox{tot}}}(s)$ and almost constant behavior in $\sigma_{\scriptsize{\mbox{el}}}(s)$ for intermediate energy range $5 \lesssim \sqrt{s} \lesssim 100$ GeV or (ii) small number of points at $\sqrt{s} > 0.5$ TeV for $pp$, $\bar{p}p$ collisions. Fig. \ref{fig:3} definitively confirms this statement by smooth increase of the $R_{\scriptsize{\mbox{e/t}}}(s)$ at $\sqrt{s} \gtrsim 5$ GeV.
Therefore the apparent effect of the constancy of $R_{\scriptsize{\mbox{e/t}}}(s)$ is dominated by the kinematic (restricted range of $\sqrt{s}$ and large gap between two subsets of experimental data) and / or methodological (small data sample) features rather than the dynamics of the interaction process. As it is known the growth of $R_{\scriptsize{\mbox{e/t}}}(s)$ is the strong argument against geometrical model and can be considered as indication on the amplification of the opacity of (anti)nucleon with the increasing of collision energy \cite{Barone-book-2002}.

%%%%%%%%%%%%%%%%%%%%%%%%%%%%%%%%%%%%%%%%%%%%%%%%%%%%%%%%%%%%%%%%%%%%%%%%
\section{Search for onset of the asymptotic region}\label{sec:4}

The analytic functions (\ref{eq:2.2}) and numerical fit results (Table \ref{tab:1}, \ref{tab:2}) allow the quantitative estimations for the ratio $R_{\scriptsize{\mbox{e/t}}}$ at various collision energies. These predictions will be within the framework of the Standard Model (SM) physics without any hypothesis for the contributions of the physics beyond the SM. On the other hand until now there is no any evidence for appearance the physic processes and / or particles beyond the SM which can be influence on the energy behavior of terms of the sets $\mathcal{G}_{pp,\bar{p}p}$. Thus the using of the approach based on the SM seems reasonable. It should be noted that continues growth of the functions (\ref{eq:2.2}) with increase of $s$ at $\sqrt{s} \geq 3$ GeV is driven by the available data. But the saturation behavior should be expected for $R_{\scriptsize{\mbox{e/t}}}(s)$ at close to the onset of the asymptotic region and at future increase of $s$. Obviously this effect can not been taken into account within model-independent fit.
As noted above the lack of experimental $\bar{p}p$ data in multi-TeV energy domain leads to the significant uncertainty for behavior of the $R_{\scriptsize{\mbox{e/t}}}(s)$ at high energies. Moreover, the most of the present facilities (RHIC, LHC) as well as the international projects for possible future higher-energy hadron collider (Future Circular Collider -- FCC-hh, Super Proton-Proton Collider -- SPPC, Very Large Hadron Collider -- VLHC) are with proton beams  \cite{FCC-arXiv-1503.09107,SPPC-arXiv-1507.03224,VLHC-Fermilab-TM-2149-2001}.
Therefore here discussion is focused on the $pp$ scattering. In addition the estimations for Facility for Antiproton and Ion Research (FAIR) and Nuclotron-based Ion Collider fAcility (NICA) energy ranges are omitted because of reasons from \cite{Okorokov-IJMPA-32-1750175-2017}.

The comparison of the various fitting curves shows that weakest dependence of the results on approximation type is at $\sqrt{s_{\scriptsize{\mbox{min}}}}=5$ GeV with preservation of the statistically acceptable $\chi^{2}/\mbox{n.d.f.}$. Therefore Table \ref{tab:5} shows the estimations for $R_{\scriptsize{\mbox{e/t}}}(s)$ calculated with help of the results for fitting functions (\ref{eq:2.2a}) and {\ref{eq:2.2b}) at $\sqrt{s_{\scriptsize{\mbox{min}}}}=5$ GeV (Table \ref{tab:1}). As expected the values of $R_{\scriptsize{\mbox{e/t}}}(s)$ coincide within errors for two functions under consideration at each $\sqrt{s}$. The ratio (\ref{eq:2.1}) reaches the asymptotic level at $\sqrt{s} \gtrsim 5-10$ PeV within 1.3--1.5 s.d. for (\ref{eq:2.2a}) and within large 1 s.d. for (\ref{eq:2.2b}) but the median values of the $R_{\scriptsize{\mbox{e/t}}}(s)$ are significantly smaller than 3/4 for both functions (\ref{eq:2.2}) so far. Thus results in Table \ref{tab:5} can be considered as indication only for the onset of the asymptotic regime in $pp$, $\bar{p}p$ elastic collisions at $\sqrt{s_{a}} \sim 5-10$ PeV. It should be noted that available experimental data allow the getting only rough estimation in order of magnitude for the energy boundary of asymptotic region. From this point of view, the estimation of $s_{a}$ agrees quantitatively with results for low-energy boundary of the asymptotic region of cross sections ($\sigma_{\footnotesize\mbox{tot}}$, $\sigma_{\footnotesize\mbox{el}}$ and $\sigma_{\footnotesize\mbox{inel}}$) obtained within approach of the partonic disks \cite{PU-58-963-2015}. Furthermore the value of the $s_{a}$ obtained here is reasonably consistent, at least, on the qualitative level with the results of the study for asymptotic behavior $\sigma_{\footnotesize\mbox{tot}}$ within the framework of AQFT approach \cite{Okorokov-PAN-81-2018} but significantly smaller than the value of $s_{a}$ which can be obtained with help of the parameter $\kappa(s)$ within Regge--eikonal model \cite{Okorokov-IJMPA-33-1850077-2018}. Thus in general numerical results of the present work confirm definitely the important conclusion from other approaches refered above, namely, the onset of the asymptotic region lies at energies larger than at least $\mathcal{O}$(100 TeV) and, more likely, even in multi-PeV energy domain. The new high-quality data at energies $\sqrt{s} \sim \mathcal{O}$(100 TeV) as well as development of phenomenological models are crucially important for improvement of the present estimations for $s_{a}$. Such data can be obtained in future collider experiments and ultra-high energy cosmic ray measurements. Perhaps, the selection of appropriate parameters which are sensitive to the transition in asymptotic regime and further simultaneous phenomenological analysis of theirs energy dependencies will allow the improvement of precision of the quantitative estimations for onset of the asymptotic region in hadronic collisions.

%%%%%%%%%%%%%%%%%%%%%%%%%%%%%%%%%%%%%%%%%%%%%%%%%%%%%%%%%%%%%%%%%%%%%%%%
\section{Summary}\label{sec:5}

The main results of this paper are the following.
Energy dependence for the ratio of elastic to total cross section $R_{\scriptsize{\mbox{e/t}}}$ is analysed
at quantitative level in details for $pp$ and $\bar{p}p$ collisions as well as for joined sample in nucleon-nucleon scattering. The analytic parameterizations with generalized logarithmic high-energy term and various functional view of low-energy term are suggested for approximation of the $R_{\scriptsize{\mbox{e/t}}}(s)$. These functions describe reasonably the all available experimental data for both $pp$ and $\bar{p}p$ with exception of the narrow region $\sqrt{s} < 2$ GeV near the low-energy boundary for the last case. The both parameterizations provide the statistically acceptable and close fit qualities for wide energy range $\sqrt{s} \geq 3$ GeV in $pp$ and $\bar{p}p$ collisions. In general the similar statements are valid for joined nucleon-nucleon data sample too but the statistically acceptable quality is observed at slightly larger low boundary of the fitted energy range $\sqrt{s_{\scriptsize{\mbox{min}}}}=5$ GeV. Nevertheless the functions suggested in the paper show the reasonable coincidence with qualitative trends in the joined dataset in almost whole experimentally available energy domain $\sqrt{s} > 2$ GeV. Thus these parameterizations allows the noticeable expansion of the described energy domain down to the lower values of $\sqrt{s}$ in comparison with other models. The ratio $R_{\scriptsize{\mbox{e/t}}}(s)$ allows the fit by constant in separate ranges at intermediate and high energies for all samples, $pp$, $\bar{p}p$ and nucleon-nucleon, under consideration but possibly the flat behavior of the $R_{\scriptsize{\mbox{e/t}}}(s)$ is explained rather by restriction of the fitted energy range and /or small number of experimental points than some dynamic features of the interaction processes. Based on the fit results the estimations are derived for the ratio $R_{\scriptsize{\mbox{e/t}}}$ in elastic $pp$ scattering at different $\sqrt{s}$ up to energy frontier $\sqrt{s}=10$ PeV. These estimations indicate that quantity under consideration reaches the its asymptotic value at $\sqrt{s} \gtrsim 5-10$ PeV within 1.0 -- 1.5 s.d. in dependence of the type of approximating function. On the other hand the large estimation uncertainties for ultra-high energies allow only the qualitative discussion for onset of the asymptotic regime for the parameter $R_{\scriptsize{\mbox{e/t}}}$ in $pp$ elastic scattering. Therefore taking into account the last statement the value $\sqrt{s_{a}} \sim 5-10$ PeV observed here can be considered only as the rough estimation for low boundary of the asymptotic energy domain. But in any way results of the paper confirm the important conclusion that the asymptotic region for hadronic interactions lies far from the energies reached in both the collider experiments and comic ray measurements.

%%%%%%%%%%%%%%%%%%%%%%%%%%%%%%%%%%%%%%%%%%%%%%%%%%%%%%%%%%%%%%%%%%%%%%%%
\section*{Acknowledgments}

The author is grateful to Prof. V. A. Petrov for valuable comments and useful
discussions. The work was supported partly by NRNU MEPhI Academic
Excellence Project (contract No 02.a03.21.0005 on 27.08.2013).

%%%%%%%%%%%%%%%%%%%%%%%%%%%%%%%%%%%%%%%%%%%%%%%%%%%%%%%%%%%%%%%%%%% FIGURES
\clearpage
%Table 1
\begin{table*}
\caption{\label{tab:1}Values of parameters for fitting of
$R_{\scriptsize{\mbox{e/t}}}(s)$ in $pp$ elastic scattering.}
\begin{center}
\begin{tabular}{ccccccc}
\hline \multicolumn{1}{c}{$\sqrt{s_{\scriptsize{\mbox{min}}}}$,} &
\multicolumn{6}{c}{Parameter} \rule{0pt}{10pt}\\
\cline{2-7}
GeV & $a_{1}$ & $a_{2}\times 10^{3}$ & $a_{3}$ & $a_{4}$ & $a_{5}$ & $\chi^{2}/\mbox{n.d.f.}$ \rule{0pt}{10pt}\\
\hline
\multicolumn{7}{c}{(\ref{eq:2.2a})} \rule{0pt}{10pt}\\
2 & $0.1596 \pm 0.0015$ & $0.136 \pm 0.016$ & $2.26 \pm 0.04$ & $1.57 \pm 0.04$   & $2.71 \pm 0.05$ & $371/87$ \\
3 & $0.016 \pm 0.009$   & $15 \pm 4$        & $0.94 \pm 0.08$ & $0.74 \pm 0.06$   & $1.20 \pm 0.10$ & $55.5/55$ \\
5 & $0.052 \pm 0.025$   & $3.8 \pm 1.6$     & $1.33 \pm 0.11$ & $0.481 \pm 0.019$ & $0.95 \pm 0.08$ & $41.9/41$ \\
\hline
\multicolumn{7}{c}{(\ref{eq:2.2b})} \rule{0pt}{10pt}\\
2 & $0.1794 \pm 0.0008$ & $(7.4 \pm 2.0)\times 10^{-5}$ & $4.79 \pm 0.09$ & $3.91 \pm 0.27$ & $1.40 \pm 0.04$ & $516/87$ \\
3 & $0.164 \pm 0.003$   & $0.06 \pm 0.04$               & $2.55 \pm 0.25$ & $0.87 \pm 0.14$ & $0.76 \pm 0.06$ & $54.5/55$ \\
5 & $0.157 \pm 0.005$   & $0.16 \pm 0.09$               & $2.23 \pm 0.19$ & $0.44 \pm 0.17$ & $0.58 \pm 0.11$ & $41.6/41$ \\
\hline
\multicolumn{7}{c}{constant} \rule{0pt}{10pt}\\
 $10^{2}$ & $0.255 \pm 0.004$ & -- & -- & -- & -- & $3.04/6$ \\
\hline
\end{tabular}
\end{center}
\end{table*}

%Table 2
\begin{table*}
\caption{\label{tab:2}Values of parameters for fitting of
$R_{\scriptsize{\mbox{e/t}}}(s)$ in $\bar{p}p$ elastic
scattering.}
\begin{center}
\begin{tabular}{ccccccc}
\hline \multicolumn{1}{c}{$\sqrt{s_{\scriptsize{\mbox{min}}}}$,} &
\multicolumn{6}{c}{Parameter} \rule{0pt}{10pt}\\
\cline{2-7}
GeV & $a_{1}$ & $a_{2}\times 10^{3}$ & $a_{3}$ & $a_{4}$ & $a_{5}$ & $\chi^{2}/\mbox{n.d.f.}$ \rule{0pt}{10pt}\\
\hline
\multicolumn{7}{c}{(\ref{eq:2.2a})} \rule{0pt}{10pt}\\
2 & $(2.50 \pm 0.04)\times 10^{-4}$ & $3.2 \pm 1.2$   & $1.52 \pm 0.13$ & $0.542 \pm 0.007$ & $0.86 \pm 0.03$ & $334/88$ \\
3 & $0.089 \pm 0.017$               & $8 \pm 4$   & $1.03 \pm 0.13$ & $0.69 \pm 0.06$   & $1.76 \pm 0.09$ & $32.3/28$ \\
5 & $0.137 \pm 0.016$                & $0.17 \pm 0.09$ & $2.33 \pm 0.20$ & $0.24 \pm 0.08$   & $1.2 \pm 0.4$   & $12.2/16$ \\
\hline
\multicolumn{7}{c}{(\ref{eq:2.2b})} \rule{0pt}{10pt}\\
2 & $0.025 \pm 0.008$ & $41 \pm 6$                    & $0.60 \pm 0.07$   & $0.776 \pm 0.024$ & $0.61 \pm 0.03$ & $266/88$ \\
3 & $0.170 \pm 0.005$ & $0.0066 \pm 0.0026$ & $3.4 \pm 0.8$     & $1.5 \pm 0.8$     & $1.17 \pm 0.22$ & $31.6/28$ \\
5 & $0.058 \pm 0.027$ & $5 \pm 3$                 & $1.331 \pm 0.024$ & $0.21 \pm 0.06$   & $0.19 \pm 0.13$ & $11.8/16$ \\
\hline
\multicolumn{7}{c}{constant} \rule{0pt}{10pt}\\
 $10^{2}$ & $0.217 \pm 0.005$ & -- & -- & -- & -- & $7.03/4$ \\
\hline
\end{tabular}
\end{center}
\end{table*}

%Table 3
\begin{table*}
\caption{\label{tab:3}Values of parameters for fitting of
$R_{\scriptsize{\mbox{e/t}}}(s)$ for joined sample for $pp$ and $\bar{p}p$ elastic
reactons.}
\begin{center}
\begin{tabular}{ccccccc}
\hline \multicolumn{1}{c}{$\sqrt{s_{\scriptsize{\mbox{min}}}}$,} &
\multicolumn{6}{c}{Parameter} \rule{0pt}{10pt}\\
\cline{2-7}
GeV & $a_{1}$ & $a_{2}\times 10^{3}$ & $a_{3}$ & $a_{4}$ & $a_{5}$ & $\chi^{2}/\mbox{n.d.f.}$ \rule{0pt}{10pt}\\
\hline
\multicolumn{7}{c}{(\ref{eq:2.2a})} \rule{0pt}{10pt}\\
2 & $(2.0 \pm 0.6)\times 10^{-4}$ & $7.8 \pm 0.4$   & $0.931 \pm 0.017$ & $0.638 \pm 0.005$ & $1.073 \pm 0.013$ & $3279/180$ \\
3 & $0.018 \pm 0.006$             & $9.9 \pm 1.5$   & $0.83 \pm 0.06$ & $0.64 \pm 0.05$  & $1.22 \pm 0.12$ & $230/88$ \\
5 & $0.153 \pm 0.004$     & $0.067 \pm 0.028$ & $2.30 \pm 0.13$ & $0.55 \pm 0.17$   & $1.92 \pm 0.24$   & $108/62$ \\
\hline
\multicolumn{7}{c}{(\ref{eq:2.2b})} \rule{0pt}{10pt}\\
2 & $0.155 \pm 0.003$ & $0.28 \pm 0.11$   & $1.80 \pm 0.12$ & $0.952 \pm 0.026$ & $0.822 \pm 0.021$ & $3191/180$ \\
3 & $0.169 \pm 0.003$ & $0.018 \pm 0.003$ & $2.7 \pm 0.3$   & $0.97 \pm 0.18$   & $0.91 \pm 0.08$ & $230/88$ \\
5 & $0.121 \pm 0.006$ & $1.0 \pm 0.3$     & $1.48 \pm 0.11$ & $0.207 \pm 0.025$ & $0.32 \pm 0.04$ & $105/62$ \\
\hline
\multicolumn{7}{c}{constant} \rule{0pt}{10pt}\\
 $10^{3}$ & $0.252 \pm 0.003$ & -- & -- & -- & -- & $9.72/9$ \\
\hline
\end{tabular}
\end{center}
\end{table*}

\clearpage
\begin{table*}
\caption{\label{tab:5} Predictions for $pp$ based on the fit results at $\sqrt{s_{\scriptsize{\mbox{min}}}}=5$ GeV.}
\begin{center}
\begin{tabular}{ccccccccc}
\hline \multicolumn{1}{l}{} &
\multicolumn{8}{c}{$\sqrt{s}$, TeV} \rule{0pt}{10pt}\\
\cline{2-9}
\multicolumn{1}{l}{Function} &
\multicolumn{3}{c}{RHIC} &
\multicolumn{2}{c}{LHC} &
\multicolumn{3}{c}{HE-LHC, LHC-ultimate} \rule{0pt}{10pt}\\
 & 0.20 & 0.41 & 0.51 & 0.90 & 14 & 28 & 33 & 42 \\
\hline
 (\ref{eq:2.2a}) & $0.19 \pm 0.05$ & $0.20 \pm 0.06$ & $0.20 \pm 0.06$ & $0.21 \pm 0.07$ & $0.27 \pm 0.11$ & $0.29 \pm 0.12$ & $0.29 \pm 0.12$ & $0.30 \pm 0.12$ \rule{0pt}{10pt}\\
 (\ref{eq:2.2b}) & $0.188 \pm 0.023$ & $0.20 \pm 0.03$ & $0.20 \pm 0.03$ & $0.21 \pm 0.04$ & $0.27 \pm 0.09$ & $0.29 \pm 0.11$ & $0.29 \pm 0.11$ & $0.30 \pm 0.12$ \\
\hline
\multicolumn{1}{l}{} &
\multicolumn{8}{c}{SPPC, FCC-hh, VLHC-I, II} \rule{0pt}{10pt}\\
 & 40 & 50 & 70.6 & 100 & 125 & 150 & 175 & 200 \\
\hline
 (\ref{eq:2.2a}) & $0.30 \pm 0.12$ & $0.30 \pm 0.13$ & $0.31 \pm 0.13$ & $0.32 \pm 0.14$ & $0.33 \pm 0.14$ & $0.33 \pm 0.15$ & $0.34 \pm 0.15$ & $0.34 \pm 0.15$ \rule{0pt}{10pt}\\
 (\ref{eq:2.2b}) & $0.30 \pm 0.12$ & $0.31 \pm 0.12$ & $0.32 \pm 0.14$ & $0.33 \pm 0.14$ & $0.34 \pm 0.15$ & $0.34 \pm 0.16$ & $0.35 \pm 0.16$ & $0.35 \pm 0.17$ \\
\hline
\multicolumn{1}{l}{} &
\multicolumn{6}{c}{ultra-high energy cosmic rays} &
\multicolumn{2}{c}{higher energies} \rule{0pt}{10pt}\\
 & 110 & 170 & 250 & 500 & 750 &$10^{3}$ & $5 \times 10^{3}$ & $10^{4}$ \\
\hline
 (\ref{eq:2.2a}) & $0.33 \pm 0.14$ & $0.34 \pm 0.15$ & $0.35 \pm 0.15$ & $0.37 \pm 0.17$ & $0.38 \pm 0.17$ & $0.39 \pm 0.18$ & $0.43 \pm 0.21$ & $0.46 \pm 0.22$ \rule{0pt}{10pt}\\
 (\ref{eq:2.2b}) & $0.33 \pm 0.15$ & $0.35 \pm 0.16$ & $0.36 \pm 0.17$ & $0.39 \pm 0.20$ & $0.40 \pm 0.21$ & $0.41 \pm 0.22$ & $0.49 \pm 0.29$ & $0.5 \pm 0.3$ \\
\hline
\end{tabular}
\end{center}
\end{table*}
%%%%%%%%%%%%%%%%%%%%%%%%%%%%%%%%%%%%%%%%%%%%%%%%%%%%%%%%%%%%%%%%%%% FIGURES
\clearpage
% Figure 1
\begin{figure}
\centering
\includegraphics[width=17.0cm,height=14.0cm]{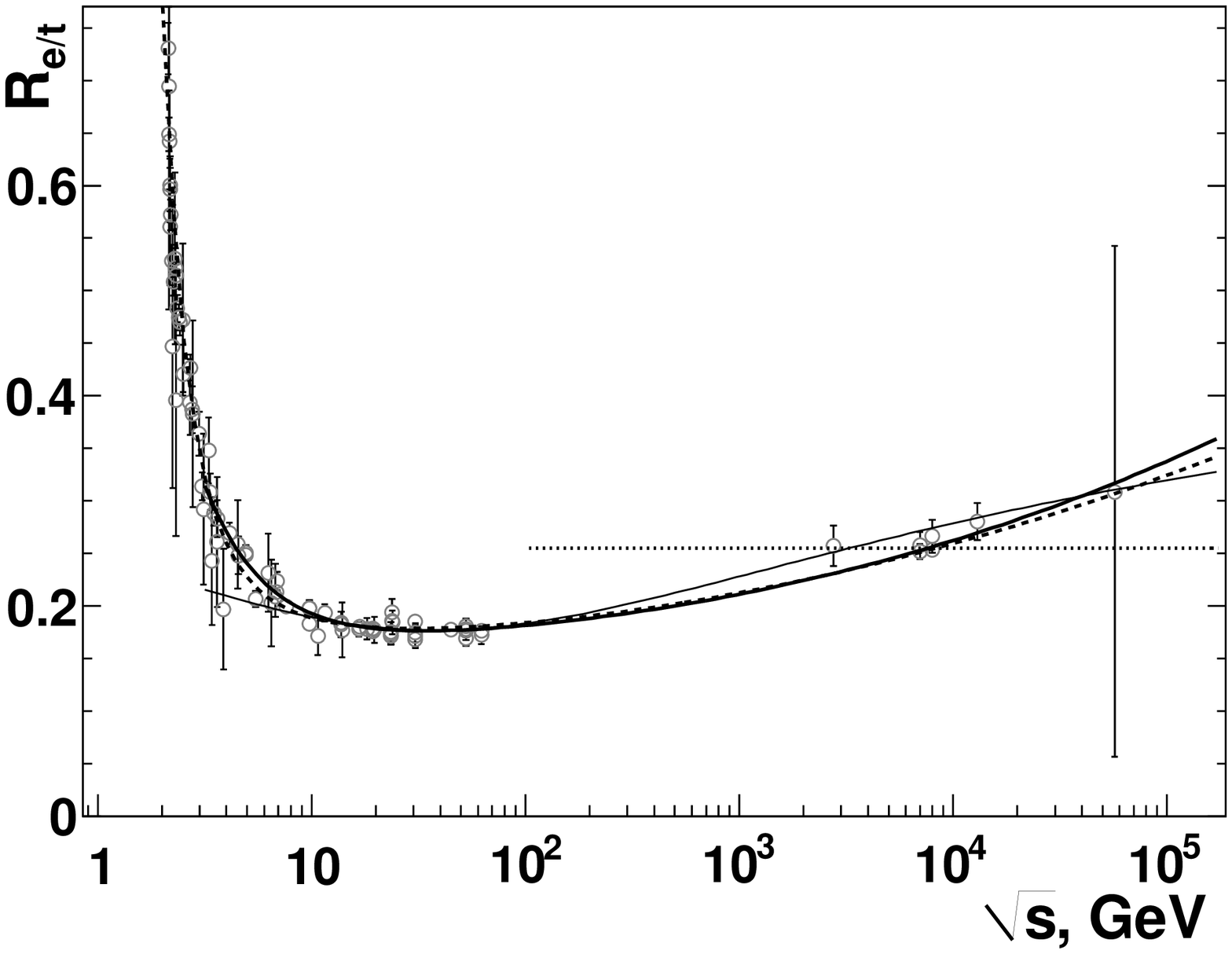}
\caption{Energy dependence for the ratio of
$\sigma_{\scriptsize{\mbox{el}}}$ to
$\sigma_{\scriptsize{\mbox{tot}}}$ in $pp$ collisions, where
dashed curve corresponds to the fit by (\ref{eq:2.2a}) at
$\sqrt{s_{\scriptsize{\mbox{min}}}}=2$ GeV, solid curve -- by (\ref{eq:2.2b}) at
$\sqrt{s_{\scriptsize{\mbox{min}}}}=3$ GeV and dotted line -- by constant at
$\sqrt{s_{\scriptsize{\mbox{min}}}}=100$ GeV (see detailed
description in the text). The thin solid line corresponds to the ratio
of the approximation for $\sigma_{\scriptsize{\mbox{el}}}(s)$ from
\cite{TOTEM-arXiv-1712.06153} to the analytic function for
$\sigma_{\scriptsize{\mbox{tot}}}(s)$ from
\cite{PDG-ChP-C40-100001-2016}. Experimental data are from
\cite{Okorokov-IJMPA-33-1850077-2018}.} \label{fig:1}
\end{figure}

% Figure 2
\begin{figure}
\centering
\includegraphics[width=17.0cm,height=14.0cm]{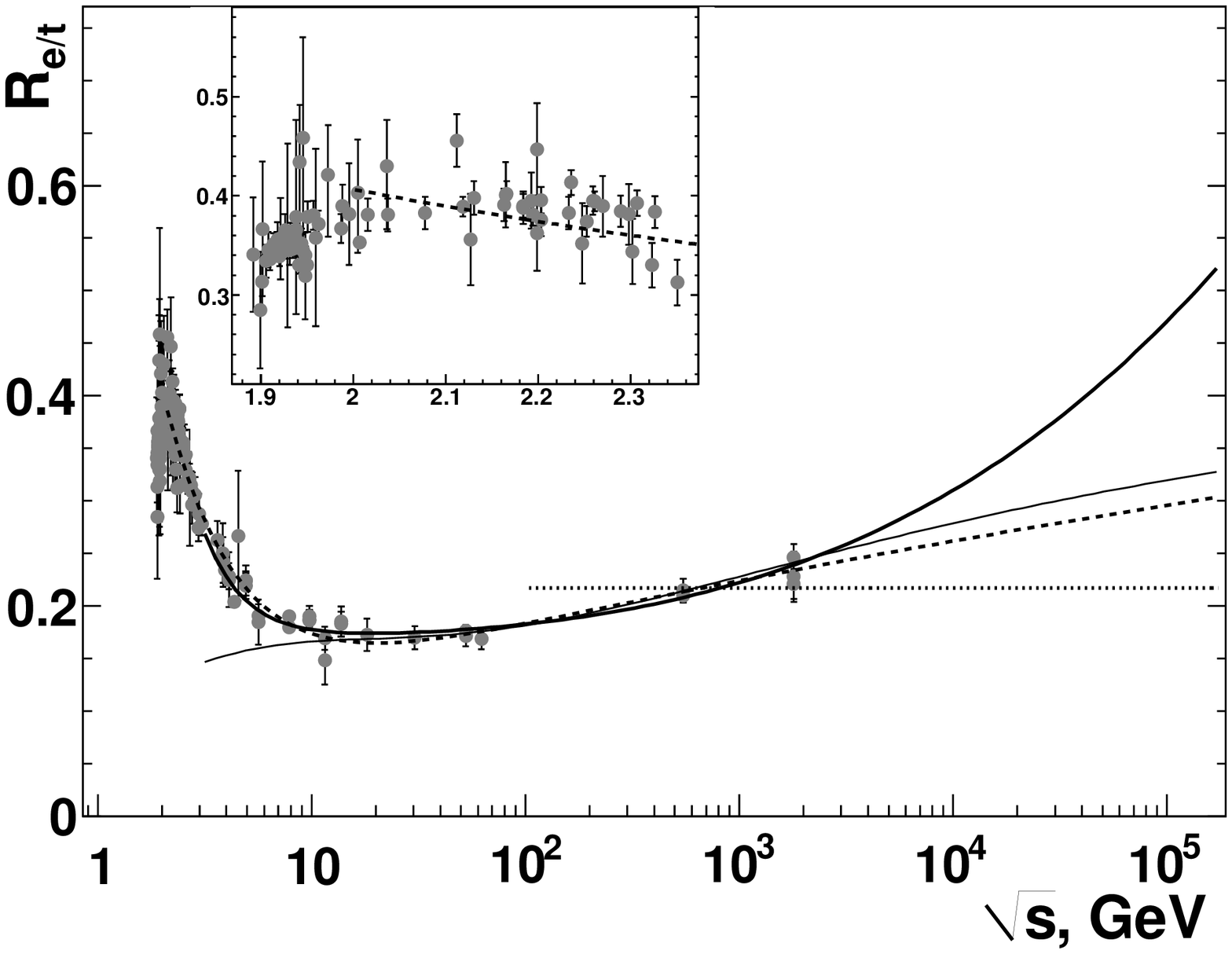}
\caption{Energy dependence for the ratio of
$\sigma_{\scriptsize{\mbox{el}}}$ to
$\sigma_{\scriptsize{\mbox{tot}}}$ in $\bar{p}p$ collisions, where
dashed curve corresponds to the fit by (\ref{eq:2.2b}) at
$\sqrt{s_{\scriptsize{\mbox{min}}}}=2$ GeV, solid curve -- by (\ref{eq:2.2b}) at
$\sqrt{s_{\scriptsize{\mbox{min}}}}=3$ GeV and dotted line -- by constant at
$\sqrt{s_{\scriptsize{\mbox{min}}}}=100$ GeV (see detailed
description in the text). The thin solid line corresponds to the ratio
of the approximation for $\sigma_{\scriptsize{\mbox{el}}}(s)$ from
\cite{TOTEM-arXiv-1712.06153} to the analytic function for
$\sigma_{\scriptsize{\mbox{tot}}}(s)$ from
\cite{PDG-ChP-C40-100001-2016}. Inner panel: the experimental points and
fit curve in the narrow range $\sqrt{s}=1.86 - 2.36$ GeV near the
low-energy boundary. Experimental data are from
\cite{Okorokov-IJMPA-32-1750175-2017}.} \label{fig:2}
\end{figure}

% Figure 3
\begin{figure}
\centering
\includegraphics[width=17.0cm,height=14.0cm]{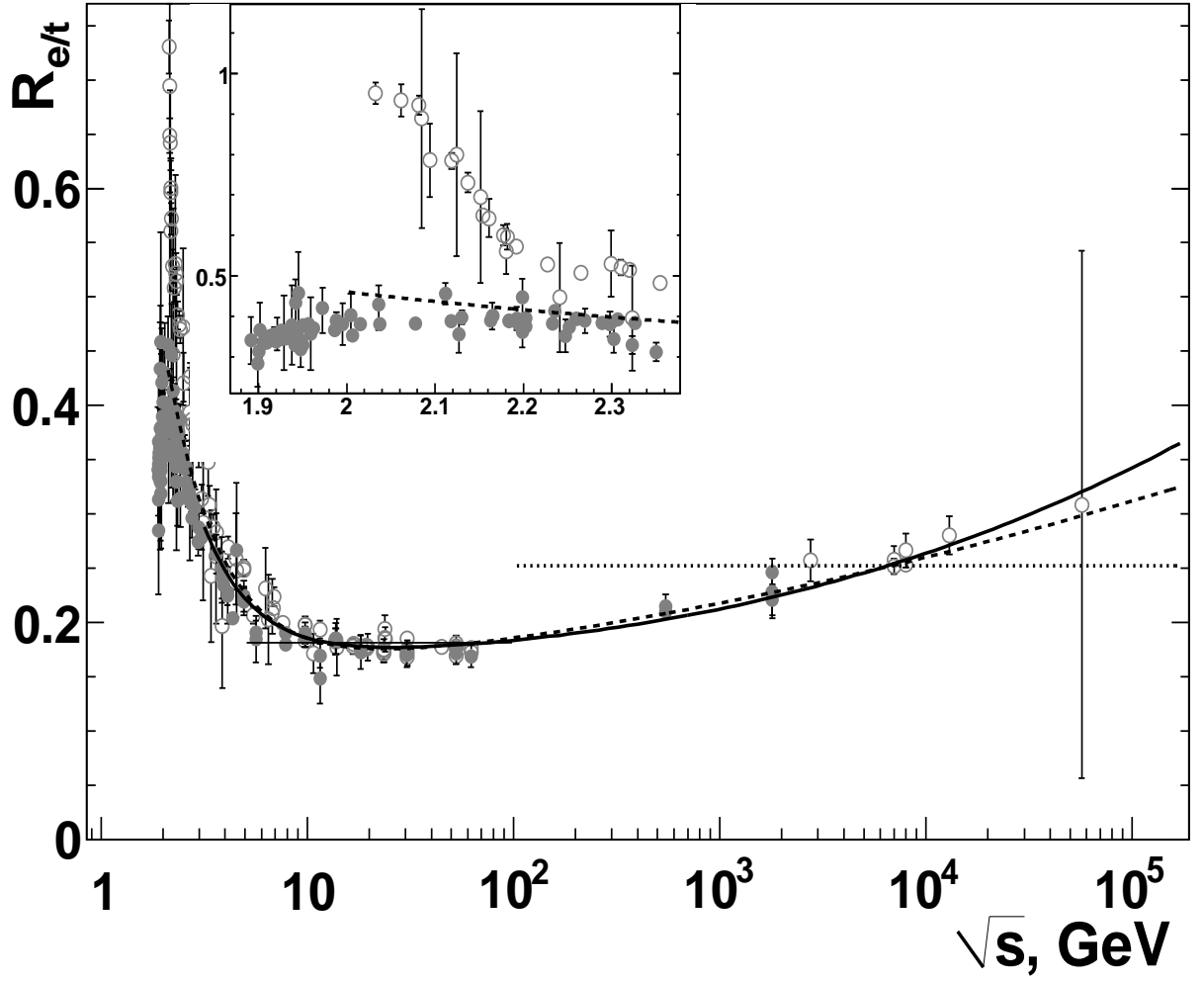}
\caption{Dependence of the ratio $R_{\scriptsize{\mbox{e/t}}}$ on
$\sqrt{s}$ for joined sample for $pp$ (open symbols) and
$\bar{p}p$ (solid symbols) collisions, where dashed curve
corresponds to the fit by (\ref{eq:2.2b}) at $\sqrt{s_{\scriptsize{\mbox{min}}}}=2$
GeV, solid curve -- by (\ref{eq:2.2b}) at $\sqrt{s_{\scriptsize{\mbox{min}}}}=3$ GeV
and dotted lines -- by constant at $\sqrt{s_{\scriptsize{\mbox{min}}}}=1$
TeV (see detailed description in the text). Thin solid curve shows
the fit by constant in the intermediate energy range at $\sqrt{s_{1}}=10$ GeV. Inner panel: the experimental points and
fit curve in the narrow range $\sqrt{s}=1.86 - 2.36$ GeV near the
low-energy boundary. Experimental data are from
\cite{Okorokov-IJMPA-33-1850077-2018,Okorokov-IJMPA-32-1750175-2017}.}
\label{fig:3}
\end{figure}

\end{document}